\documentclass[twocolumn,aps,showpacs,prl]{revtex4}
\usepackage{graphicx} 
\begin{document}
\title{Structural and electronic properties of the interface between the high-k oxide LaAlO$_\mathbf{3}$ and Si(001)}
\author{Clemens J. F\"orst,$^{1,2,3}$ 
        Karlheinz Schwarz,$^{2}$  and
        Peter E. Bl\"ochl$^{1}$}
\affiliation{$^1$ Clausthal University of Technology, Institute for
Theoretical Physics, Leibnizstr.10, D-38678 Clausthal-Zellerfeld,
Germany}
\affiliation{$^2$ Vienna University of Technology, Institute for
Materials Chemistry, Getreidemarkt 9/165-TC, A-1060 Vienna, Austria}
\affiliation{$^3$ Departments of Nuclear Science and Engineering and
 Materials Science and Engineering, Massachusetts Institute of Technology, Cambridge,
Massachusetts 02139, USA}
\date{\today}
\begin{abstract}

The structural and electronic properties of the LaAlO$_3$/Si(001)
interface are determined using state-of-the-art electronic structure
calculations.  The atomic structure differs from previous proposals, but
is reminiscent of La adsorption structures on silicon. A phase diagram
of the interface stability is calculated as a function of oxygen and Al
chemical potentials.  We find that an electronically saturated interface
is obtained only if Al atoms substitute some of the interfacial Si
atoms.  These findings raise serious doubts whether LaAlO$_3$ can be
used as an epitaxial gate dielectric.

\end{abstract}
\maketitle


One of the key challenges of semiconductor technology in the coming
years is the replacement of SiO$_2$ as a gate dielectric in
microelectronic devices~\cite{roadmap}. Due to the downscaling of device
dimensions, conventional gate oxides reach a thickness where they lose
their insulating properties due to tunneling currents.  The replacement
of SiO$_2$ with \mbox{high-k} oxides, having a larger dielectric
constant (k), allows to increase the capacitance of the gate stack with
a gate oxide of greater physical thickness.  However, the growth of
high-quality interfaces between Si and \mbox{high-k} oxides remains
challenging.

While -- in a first phase -- amorphous oxides based on ZrO$_2$ or
HfO$_2$ will be likely employed, industry requires solutions for
crystalline oxides with an epitaxial interface to the silicon substrate
as soon 2013~\cite{roadmap}.  The latest edition of the International
Roadmap for Semiconductors~\cite{roadmap} lists the perovskite LaAlO$_3$
(LAO) as a candidate epitaxial oxide.  So far, however, there are few
demonstrations of atomically well defined interfaces of perovskite
oxides grown on silicon, namely SrTiO$_3$ (STO) and BaTiO$_3$, as shown
by McKee and coworkers~\cite{McKee98,Hu03}.  The main reason why STO is not
considered as a viable gate dielectric is that its electron injection
barrier in the structures prepared so far is too small for transistor
applications~\cite{Chambers01}.  Nevertheless, ab-initio
calculations~\cite{Ashman04,Foerst04} provide strong evidence that the
band offset can be engineered to meet technological requirements.
Besides perovskite oxides, films with fluorite derived structures are
considered promising candidates for epitaxial high-k oxides on
silicon~\cite{Guha02,Seo03}.

LAO has a dielectric constant of 24~\cite{Schwab97}, a bandgap of
5.5\,eV~\cite{Lim02} as compared to 3.2\,eV for STO and a large
conduction band offset with silicon in its amorphous
phase~\cite{Edge04}. Thus it seems to provide excellent properties for
use as a gate oxide. So far, however, there has been no report of
heteroepitaxial growth of LAO on silicon.  Experimental studies have
been restricted to amorphous or polycrystalline LAO layers on
silicon~\cite{Park03,Li03,Edge04a,Edge04}.  Theoretical contributions
are limited to a study of a $(2\times 1)$ reconstructed
interface~\cite{Robertson04}. Recently, epitaxial growth of silicon with
an atomically well defined interface on LAO has been
reported~\cite{Klenov05}. While the reverse growth condition is required
for \mbox{high-k} gate oxide applications, a first understanding of the
interface was obtained by atomic resolution high-angle annular dark
field imaging (HAADF) in scanning transmission electron microscopy
(STEM).

Motivated by the work of Klenov et al.~\cite{Klenov05}, we investigated
the interface between LAO and silicon with state-of-the-art
first-principles calculations. We provide a detailed structural model
for the interface and discuss its chemical phase stability with respect
to oxygen stoichiometry and doping by Al. We demonstrate that doping is
a prerequisite for obtaining an electronically saturated interface as
required for its use in a gate stack.


Our results are based on density functional
theory calculations~\cite{Kohn,KohnSham,PBE} using the projector
augmented wave method~\cite{Blo94,Blo03} combined with the ab-initio
molecular dynamics scheme~\cite{Car85}.  All calculations were performed in
$(3\times 2)$ silicon surface unit cells with a slab thickness of five
atomic layers of silicon and 2.5 unit cells (5 atomic layers) of LAO.
The LaO layer terminating the LaAlO$_3$ slab has been alloyed with
50\,\%\ Sr to satisfy the electron count at the interface to the vacuum
region of around 8\,\AA~\cite{Tasker79}. We have used a regular mesh
corresponding to 36 k-points per  $(1\times 1)$ surface unit cell.
Further details about the computational procedure can be found in our
previous publications in this
field~\cite{Foerst04,Ashman04,Ashman04a,Foerst05}.


\begin{figure}
\centering 
\includegraphics[width=7cm]{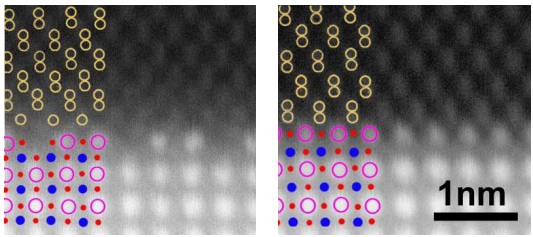}
\caption{(Color online) HAADF-STEM images of the interface
between LAO and Si~\cite{Klenov05}. For color coding see
Fig.~\ref{fig:strcs}.}
\label{fig:tem}
\end{figure}

The experimental evidence contained in the scanning transmission
electron microscopy images  effectively narrows the phase space to be
considered in our search for the interface structure.  Analysis of these
images, shown in Fig.~\ref{fig:tem}, reveals (1) that the La atoms at
the interface exhibit a $(3\times 1)$ reconstruction, (2) that there are
2/3~monolayer (ML) of La at the interface with double rows of La atoms
separated by La-vacancy rows and (3) that the first AlO$_2$ layer is a
full bulk layer as discussed by Klenov et al.~\cite{Klenov05}. Other
interface configurations, such as one shifted by one half of a
unit-cell, were observed as well. They may, however, be related to steps
of the LAO substrate. 

Despite the high quality of the HAADF-STEM images certain details of the
interface structure cannot be resolved unambiguously.  In particular,
the details of the Si network at the interface and the oxygen content of
the interface cannot be determined from these images.  These questions
will be addressed first in this Letter.  Later we will also investigate the
effects of Al doping.

\begin{figure} \centering
\includegraphics[angle=0,width=7cm]{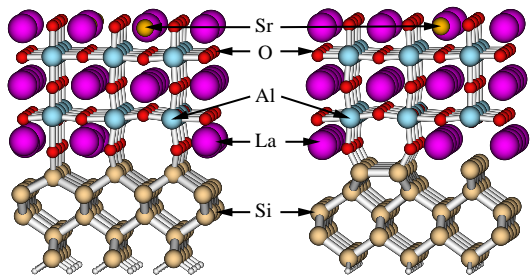}
\caption{(Color online) The two limiting cases for the interface structure. Left panel:
type $A$ where the dimer rows are orthogonal to the missing La rows; right
panel: type $B$ where the dimer rows are parallel to the missing La row.
Within each structure type the oxygen content at the interface can vary.}
\label{fig:strcs} 
\end{figure}

The first question to be addressed is that of the relative orientation
of the La-vacancy-rows relative to the silicon dimer rows. The two
different choices lead to the structure types denoted $A$ and $B$ as
shown in Fig.~\ref{fig:strcs}. In structure $A$ the La-vacancy-rows are
oriented orthogonal to the Si dimer rows and in structure $B$ they are
parallel.  Structure type $B$ reflects the stable surface structures of
La adsorbed at silicon at a coverage of 2/3 ML, as obtained by
first-principles calculations~\cite{Ashman04a}.  We furthermore
considered interfaces $C$ and $D$ which result from interface $B$ after
removal of the silicon dimer rows and isolated silicon atoms,
respectively.  Removing both silicon dimers and undimerized silicon
atoms from structure $B$ yields interface $A$.  Thus we considered a
consistent set of trial structures in our search, which had already been
identified by Klenov et al.~\cite{Klenov05}. 

The stability of the interface structures has to be considered in the light of oxygen
stoichiometry, because the interface can exchange oxygen with the oxide by
creating or annihilating oxygen vacancies. Following our observation that
oxygen attacks the silicon dangling bonds at the STO/Si
interface~\cite{Foerst04}, we investigated variants of the interfaces types
$A-D$ which are obtained by varying the number of oxidized Si dangling
bonds.

\begin{figure}
\centering
\includegraphics[width=7cm,clip=true]{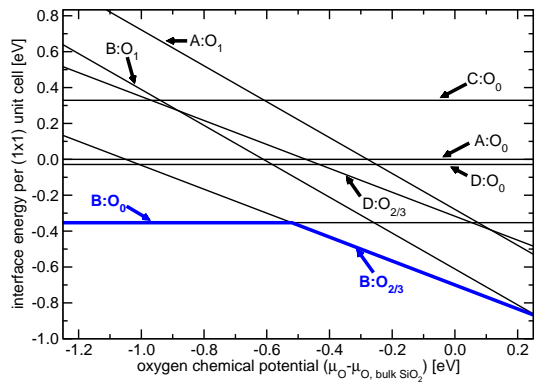}
\caption{(Color online) Interface energies per $(1\times1)$ unit cell for the interface types
$A-D$ as a function of oxygen chemical potential relative to $A$:O$_0$.  The zero for the oxygen
chemical potential is the stability line of Si and SiO$_2$. The labels
indicate the interface type and the oxygen content per $(1\times 1)$
surface unit cell denoted by $x$ in ``O$_x$". Interfaces of type $B$ are stable for the entire
range of oxygen chemical potentials.}
\label{fig:1dphase}
\end{figure}

For the entire range of oxygen stoichiometries we find that variants of
structure type $B$ are more stable than those of interfaces $A,C,D$ by
at least 0.33\,eV per $(1\times 1)$ unit cell of the Si surface.  In
Fig.~\ref{fig:1dphase} the energies are shown as function of oxygen
chemical potential, which reflects the oxygen partial pressure.
Oxidation of the interface starts above an oxygen chemical potential of
$-0.52$\,eV.  The first oxygen atoms attack the dangling bonds of the
Si-dimers, while the isolated silicon atoms of structure type $B$
underneath the La double-row remain vacant.  The undimerized silicon
atoms resist oxidation up to at a chemical potential of $+0.27$\,eV,
which lies beyond the onset of bulk-silicon oxidation.  

For use in device applications the interface must be electrically
inactive. This implies that the Fermi level is not pinned at the
interface. The Fermi level can be pinned either by the interface states
located in the band gap of silicon or by the interface being charged so
that the charge compensating conduction electrons or holes pin the Fermi
level either in the conduction band or the valence band of silicon.

We find that there are no deep interface states in the band gap of
silicon, but that all interfaces considered so far are charged.  Due to
the importance of this finding we discuss the electron count in
some detail:

Let us first investigate the oxide: Like STO, LAO crystallizes in the
perovskite structure, which consists of alternating layers LaO and
AlO$_2$ stacked in (001) direction.  Using the formal charges La$^{3+}$,
Al$^{3+}$ and O$^{2-}$, the layers alternate in being positively charged
(LaO)$^{+}$ and negatively charged (AlO$_2$)$^{-}$. This differs from
STO where both layers are charge neutral.  This seemingly subtle
difference does have quite significant implications on the atomic and
electronic structure of the interface.  In the bulk oxide half of the
charge of an AlO$_2$ layer is compensated by the LaO layer above and the
other half by the one below~\cite{Tasker79}. Thus, an AlO$_2$ terminated
surface of LAO has a surplus of $\frac{1}{2}$ electron per ($1\times1$)
unit cell.

Let us now turn to silicon: Compared to the Si surface, the dangling
bond states of the Si atoms at the interface to an oxide are shifted
down, almost into the valence band of silicon due to the electrostatic
attraction by nearby cations~\cite{Ashman04,Ashman04a}. If the Fermi level is in
the band-gap, these orbitals are filled and the silicon atoms have a
negative charge. If an oxygen atom binds to silicon, a bonding orbital
of mostly oxygen character is formed in the valence band, and an
anti-bonding orbital of mostly silicon character is shifted up into the
conduction band, resulting in a (formally) positively charged silicon
atom.  Thus, the silicon dangling bond is amphoteric in that it changes
its formal charge state from negative to positive, while two electrons
are transfered to the oxygen atom. For a three-fold coordinated silicon
atom we obtain 
\begin{center} 
Si$^-$+O$^{0}\leftrightarrow$ Si$^+$O$^{2-}$ 
\end{center} 
while for a two-fold coordinated silicon atom we obtain 
\begin{center} 
Si$^{2-}$+2O$^{0} \leftrightarrow$ Si$^0$O$^{2-}$+O$^{0} \leftrightarrow$ Si$^{2+}$O$^{2-}_2$ 
\end{center}

This explains the somewhat counterintuitive observation that introducing
oxygen atoms bound to the silicon atoms at the interface does not affect
the net charge of the interface.

Having determined the formal charges of the two half-crystals, we can
simply count the remaining charges to determine the net charge of the
interface.  The AlO$_2$ layer contributes a charge of $-\frac{1}{2}$~e
per $(1\times 1)$ unit cell.  The La$^{3+}_{\frac{2}{3}}$ layer at the
interface contributes a charge of $+2$~e. Since the oxygen stoichiometry
of the interface does not change the electron count, we consider the
corresponding unoxidized interfaces $A-D$.  Interfaces $A,D$ have one
Si-dangling bond per $(1\times 1)$ unit cell while interfaces $B,C$ have
$\frac{4}{3}$ dangling bonds.  Since each dangling bond contributes with
$-1$~e, we obtain charges $q_A=q_D=\frac{1}{2}$~e for interfaces $A$ and
$D$, while interface $B$ and $C$ contribute with
$q_B=q_C=\frac{1}{6}$~e.

To obey the requirement of charge neutrality, this charge must be
compensated by the corresponding number of conduction electrons in
silicon, which pins the Fermi level at the conduction band for all
interfaces considered. Thus none of the interfaces discussed so far is
suitable for use in a gate stack.


To overcome the above mentioned deficiency, holes have to be
introduced.  This can be achieved by doping either the oxide or silicon
with an acceptor. The only reasonable choice is to replace silicon
substitutionally by Al atoms, which act as acceptors.  

To reduce the high-dimensional phase space we limited our search to the
structure types $A$ and $B$ as the most stable representatives for the
two interface charges 1/6 and 1/2.  For both structure types we
considered structures with the dimer dangling bonds either being
unoxidized, denoted as $A/B$:O$_0$, or fully oxidized, denoted as
$A$:O$_1$ and $B$:O$_{2/3}$, since these are the oxygen contents that
are stable in the thermodynamically accessible range of oxygen chemical
potentials (compare Fig.~\ref{fig:1dphase}).  For structure
$B$:O$_{2/3}$ we found that replacing the divalent silicon atom by Al is
energetically highly unfavorable compared to replacing one of the
Si-dimer atoms. Therefore, we limited our search to substitutions of Si
dimer atoms. For the remaining structures we substituted one to four
silicon atoms by Al.

\begin{figure}
\centering
\includegraphics[width=7cm,clip=true]{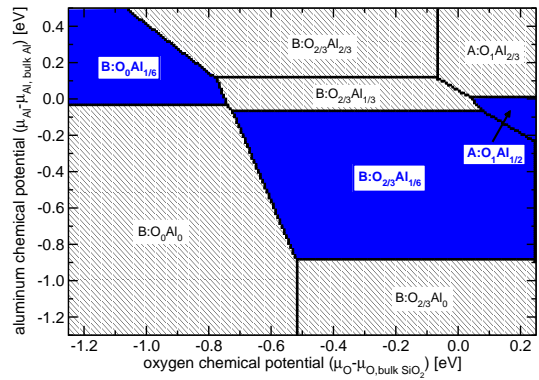}
\caption{(Color online) Interface stability as a function of oxygen and aluminum
chemical potentials. The interfaces with unpinned Fermi level are shaded in dark
(blue). The chemical potentials are plotted relative to the
stability line of Si and SiO$_2$ as well as to the chemical potential in
bulk Al.  Labels as in Fig.~\ref{fig:1dphase}}
\label{fig:2dphase}
\end{figure}

The results are composed in the phase diagram shown in
Fig.~\ref{fig:2dphase}.  The chemical potentials are given relative to
the equilibrium between bulk Si and SiO$_2$ for oxygen and to a
reservoir of metallic Al on the other hand.  Thus only the portion with
negative chemical potentials of both, O and Al, is thermodynamically
accessible.  

To compensate the charge of interface types $A$ and $B$ respectively,
substitutional doping with $\frac{1}{2}$ and $\frac{1}{6}$~ML of Al is
required to obtain an electrically inactive interface. Three such
regions in the phase diagram are found: $B$:O$_0$Al$_{1/6}$,
$B$:O$_{2/3}$Al$_{1/6}$ and $A$:O$_1$Al$_{1/2}$. 

\begin{figure}
\centering
\includegraphics[width=6cm,clip=true]{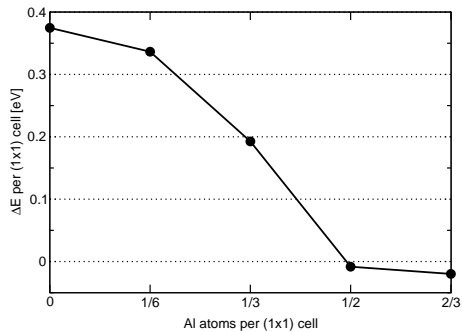}
\caption{The energy difference between the interfaces $A$:O$_1$ and
  $B$:O$_\frac{2}{3}$ shown in Fig.~\ref{fig:strcs} as a function of $N_{Al}$,
  the number of substitutional Al atoms at the interface per $(1\times1)$ unit
  cell. $\mu_\mathrm{O}$ has been set equal to the coexistance of
  Si and SiO$_2$.}
\label{fig:de_vs_nal}
\end{figure}

While interface $B$ is clearly more favorable for low Al-stoichiometry
below $\frac{1}{6}$ Al per ($1\times1$) unit cell, both interface types
are nearly isoenergetic for large Al-contents, that is for more than
$\frac{1}{2}$~Al per unit cell. This can be rationalized by the
dependence of the formation energy of a substitutional Al in silicon as
function of the electron chemical potential. The formation energy
depends on the electron chemical potential since
$E^f(\mu)=E^f(\mu=\epsilon_v)-\mu$ for values within the silicon band
gap, i.e. for $\epsilon_v<\mu<\epsilon_c$. Thus it costs less energy to
introduce an Al atom, when the Fermi level is pinned at the conduction
band than at the valence band.  This leads to the step-like shape of the
relative energies of structure type $A$ and $B$ as a function of the Al
content shown in Fig.~\ref{fig:de_vs_nal}.  If the Fermi levels are
equal, to a first approximation, the energy to substitute one Si by an
Al atom is identical for both interfaces, explaining the approximately
flat behavior for $N_{Al}<\frac{1}{6}$ and for $N_{Al}>\frac{1}{2}$. In
the crossover region, that is for $\frac{1}{6}<N_{Al}<\frac{1}{2}$ the
Fermi level for the $A$-type interface is pinned at the conduction band
while that of the $B$-type interface is pinned at the valence band,
resulting in a slope of approximately the silicon band gap versus
$N_{Al}$.  We may speculate that the strong dependence of the
interfacial energy on the Al content at the interface provides a
thermodynamical driving force towards saturated interfaces.

Furthermore, Fig.~\ref{fig:2dphase} reveals that oxidation of the
dangling bonds significantly stabilizes the substitution of Si by Al,
since the corresponding phase boundaries are shifted to lower Al
chemical potentials when the dangling bonds are oxidized.

In conclusion, we determined the atomic structure and stoichiometry of
the LaAlO$_3$/Si interface. Without doping, the Fermi level is pinned in
the conduction band of silicon, a result that is independent of the
oxygen content of the interface. For a technologically viable interface,
a specified amount of interfacial Si atoms needs to be replaced by Al. A
phase diagram for variable oxygen- and aluminum-content is provided. Our
results cast doubt that LaAlO$_3$ can be used as epitaxial gate oxide.
Its use depends on the ability to precisely control the aluminum content
of the interface which may, however, be facilitated by a
self-compensation mechanism.

The authors thank D.~Klenov and S.~Stemmer for providing us with the
STEM images prior to publication as well as for stimulating discussions.
We furthermore thank A.~Reyes Huamantinco for contributions in the
inital stage of the project. C.F. acknowledges the support of S.~Yip and
J.~Li.  This project received funding from the European Commission
(project ``ET4US"), HLRN, the Austrian Science Fund (project ``AURORA"),
NSF (CHE-0434567, IMR-0414849) and SRC.


%

\begin{thebibliography}{99}
%
\bibitem{roadmap} International Technology Roadmap for Semiconductors,
2003 Ed. http://public.itrs.net/
%
\bibitem{McKee98} R.A. McKee, F.J. Walker and M.F. Chisholm, Phys. Rev.
Lett. \textbf{81}, 3014 (1998).
%
\bibitem{Hu03} X.~Hu et al., Appl.\,Phys.\,Lett. \textbf{82}, 203
(2003).
%
\bibitem{Chambers01} S.A.~Chambers,  Y. Liang, Z. Yu, R. Droopad, and J. Ramdani,
J.\,Vac.\,Sci.\,Technol.\,A \textbf{19}, 934 (2001).
%
\bibitem{Ashman04} C.R. Ashman, C.J. F\"orst, K. Schwarz and P.E.
Bl\"ochl, Phys. Rev. B \textbf{69}, 75309 (2004).
%
\bibitem{Foerst04} C.J. F\"orst, C.R. Ashman, K. Schwarz and P.E.
Bl\"ochl, Nature \textbf{427}, 53 (2004).
%
\bibitem{Guha02} S. Guha, N.A. Bojarczuk, and V. Narayanan, Appl.\,Phys.\,Lett.
\textbf{80}, 766 (2002)
%
\bibitem{Seo03} J.W. Seo, J. Fompeyrine, A. Guiller, G. Norga, C.
Marchiori, H. Siegwart, and J.P. Locquet, Appl.\,Phys.\,Lett.
\textbf{83}, 5211 (2003)
%
\bibitem{Schwab97} R. Schwab, R. Sporl, P. Severloh, R. Heidinger, and J. Halbritter, in
Applied Superconductivity 1997, Vols 1 and 2 (1997), p. 61.
%
\bibitem{Lim02} S.G. Lim, S. Kriventsov, T.N. Jackson, J.H. Haeni, D.G. Schlom, A.M.
Balbashov, R. Uecker, P. Reiche, J.L. Freeouf, and G. Lucovsky,
J.\,Appl.\,Phys.  \textbf{91}, 4500 (2002).
%
\bibitem{Edge04} L.F. Edge, D.G. Schlom, S.A. Chambers, E. Cicerrella,
J.L. Freeouf, B. Hollander and J. Schubert, Appl.\,Phys.\,Lett.
\textbf{84}, 726 (2004)
%
\bibitem{Park03} B.E. Park and H. Ishiwara
Appl.\,Phys.\,Lett.
\textbf{82}, 1197 (2003)
%
\bibitem{Li03} A.D. Li, Q.Y. Shao, H.Q. Ling, J.B. cheng, D. Wu, Z.G.
Liu, N.B. Ming, C. Wang, H.W. Zhou and B.Y. Nguyen, Appl.\,Phys.\,Lett.
\textbf{83}, 3540 (2003)
%
\bibitem{Edge04a} L.F. Edge, D.G. Schlom. R.T. Brewer, Y.J. chabal, J.R.
Williams, S.A. Chambers, C. Hinkle, G. Lucovsky, Y. Yang, S. Stemmer, M.
Coplel, B. Holl\"ander and J. Schubert
Appl.\,Phys.\,Lett.
\textbf{84}, 4629 (2004)
%
\bibitem{Robertson04} J. Robertson and P. W. Peacock,
Mat.\,Res.\,Soc.\,Symp.\,Proc.  \textbf{786}, 23 (2004),
Phys,\,Stat.\,Sol.~B \textbf{241}, 2236 (2004).
%
\bibitem{Klenov05} Dmitri O. Klenov, Darrel G. Schlom, Hao Li and
Susanne Stemmer, Jap.\,J.\,Appl.\,Phys. \textbf{44}, L617 (2005)
%
\bibitem{Kohn} P.~Hohenberg and W.~Kohn, Phys. Rev. \textbf{136}, B864
(1964).
%
\bibitem{KohnSham}W. Kohn and L.J.~Sham, Phys. Rev. \textbf{140}, A1133
(1965).
%
\bibitem{PBE}J.P.~Perdew, K.~Burke, and M.~Ernzerhof, Phys.~Rev.~Lett.
\textbf{77}, 3865 (1996).
%
\bibitem{Blo94}P.E. Bl\"ochl, Phys. Rev. B \textbf{50}, 17953 (1994).
%
\bibitem{Blo03} Peter E. Bl\"ochl, Clemens J. F\"orst and Johannes
Schimpl, Bull. Mater. Sci. \textbf{26}, 33 (2003)
%
\bibitem{Car85}R. Car and M. Parrinello, Phys. Rev. Lett. \textbf{55},
2471 (1985).
%
\bibitem{Tasker79} P.W. Tasker, J.\,Phys.\,C \textbf{12}, 4977 (1979)
%
\bibitem{Ashman04a} C.R. Ashman, C.J. F\"orst, K. Schwarz and P.E.
Bl\"ochl, Phys. Rev. B \textbf{70}, 155330 (2004).
%
\bibitem{Foerst05} C.J. F\"orst, C.R. Ashman, K. Schwarz and P.E.
Bl\"ochl, Microelectronic Engineering \textbf{80}, 402 (2005).
%
\end{thebibliography}
\end{document}